\documentclass[prl,twocolumn,showpacs,floatfix,superscriptaddress]{revtex4-1}
\usepackage{amsmath, graphicx}

\begin{document}

\title{Point defects on graphene on metals}

\author{ M. M. Ugeda }
\affiliation{D. de F\'{\i}sica de la Materia Condensada,Universidad Aut\'onoma de Madrid, E-28049 Madrid, Spain}
\author{ D. Fern\'andez-Torre }
\affiliation{ D. de F\'{\i}sica Te\'orica de la Materia Condensada,Universidad Aut\'onoma de Madrid, E-28049 Madrid, Spain }
\author{I. Brihuega}
\email[Corresponding author.\\Email address: ]{ivan.brihuega@uam.es}
\affiliation{D. de F\'{\i}sica de la Materia Condensada,Universidad Aut\'onoma de Madrid, E-28049 Madrid, Spain}
\author{ P. Pou}
\affiliation{ D. de F\'{\i}sica Te\'orica de la Materia Condensada,Universidad Aut\'onoma de Madrid, E-28049 Madrid, Spain }
\author{ A.J. Mart\'{\i}nez-Galera }
\affiliation{D. de F\'{\i}sica de la Materia Condensada,Universidad Aut\'onoma de Madrid, E-28049 Madrid, Spain}
\author{ Rub\'en P\'erez }
\affiliation{ D. de F\'{\i}sica Te\'orica de la Materia Condensada,Universidad Aut\'onoma de Madrid, E-28049 Madrid, Spain }
\author{J. M. G\'omez-Rodr\'{\i}guez}
\affiliation{D. de F\'{\i}sica de la Materia Condensada,Universidad Aut\'onoma de Madrid, E-28049 Madrid, Spain}

\begin{abstract}

Understanding the coupling of graphene with its local environment is critical to be able to integrate it in tomorrow's electronic devices. Here we show how the presence of a metallic substrate affects the properties of an atomically tailored graphene layer. We have deliberately introduced single carbon vacancies on a graphene monolayer grown on a Pt(111) surface and investigated its impact in the electronic, structural and magnetic properties of the graphene layer. Our low temperature scanning tunneling microscopy studies, complemented by density functional theory, show the existence of a broad electronic resonance above the Fermi energy associated with the vacancies. Vacancy sites become reactive leading to an increase of the coupling between the graphene layer and the metal substrate at these points; this gives rise to a rapid decay of the localized state and the quenching of the magnetic moment associated with carbon vacancies in free-standing graphene layers.

\end{abstract}


\maketitle

Exciting properties were supposed for graphene since long time ago \cite{Wallace'47}. However, it was not till 2004 \cite{Novoselov'04} that graphene ceased being a theoretical chimera to become the object of desire of the scientific community. In just few years, most of these extraordinary properties have already been demonstrated \cite{Novoselov'05b,Zhang'05, Miller'09} and many others are emerging  as a result of the tremendous experimental and theoretical efforts devoted to this material \cite{Castro Neto'09, Geim'09}. As a consequence, graphene has undoubtedly become one of the most promising candidates to play a key role in future technology. Many of the experimental efforts have been invested in growing larger and higher quality graphene layers and also in understanding and controlling the coupling of graphene with other materials, in particular with metals, a must to incorporate graphene to real devices. Epitaxial graphene on metals represents an ideal route to fulfill both requirements. Highly perfect graphene sheets can be grown on various metals \cite{N'Diaye'06, de Parga'08, Wintterlin'09} and very recently macroscopic-sized graphene films have been grown and subsequently transferred to arbitrary substrates \cite{Kim'09, Li'09}. The interaction of graphene layers with the different metallic substrates strongly depends on the metal itself \cite{Giovannetti'08, Preobrajenski'08}, but according to its strength two main groups can be identified: strongly and weakly interacting systems \cite{Wintterlin'09}. While properties of graphene monolayers belonging to the first group can be quite different with respect to the free-standing case \cite{Gruneis'08}, in the weakly interacting systems the electronic structure of ideal graphene is basically preserved, as revealed by the experimental observation of Dirac cones comparable to those of perfect graphene \cite{Pletikosic'09, Sutter'09}. 

Till date, graphene on metals research has basically focused on pristine graphene monolayers adsorbed on metal surfaces, which are at present quite well understood \cite{Wintterlin'09, Pletikosic'09, Sutter'09}. In contrast, despite the pressing challenge to tailor graphene layers in order to fully exploit graphene's potential, very little is known about the influence of the metallic substrate in the properties of modified graphene layers \cite{Balog'10}. Here, we show that while properties of pristine graphene adsorbed on a weakly interacting metal like Pt(111) are reasonably preserved, the situation is dramatically different when point defects are introduced in the graphene layer. 

Our starting point is a perfectly clean graphene monolayer adsorbed on Pt(111), one of the weakest interacting graphene-metal systems \cite{Wintterlin'09, Sutter'09}. As a result of such a weak interaction, the graphene layer presents several orientations with respect to the Pt(111) surface, giving rise to various moir\'e  patterns \cite{Land'92,Sutter'09, Enachescu'99, Otero'10}. In addition, recent photoemission experiments have shown the existence of ${\pi}$ bands presenting linear dispersion, the same Fermi velocity as in free standing graphene (FSG) and a Dirac point slightly shifted to $\sim$ +300 meV \cite{Sutter'09}. We formed a complete graphene monolayer on the clean Pt(111) surface by chemical vapor deposition of ethylene in ultrahigh vacuum (UHV) environments at temperatures above 1275 K. Our experiments were performed at 6 K using a home-made low temperature scanning tunneling microscope (LT-STM) operating in UHV \cite{Miguel'11}. Figure 1a shows a large STM image of an atomically perfect graphene/Pt(111) surface, where two moir\'e patterns can be identified. In the upper-left corner an $(\sqrt{21}\times\sqrt{21})R11^\circ$ moir\'e  is found, while the rest of the image shows a 3$\times$3 moir\'e (both periodicities with respect to the graphene lattice). Such a 3$\times$3 superperiodicity is due to a 19.1$^\circ$ angle between graphene and the Pt(111) surface, corresponding with a lattice misfit of 0.60\%  \cite{Enachescu'99}.

\begin{figure}
\includegraphics[width=85mm]{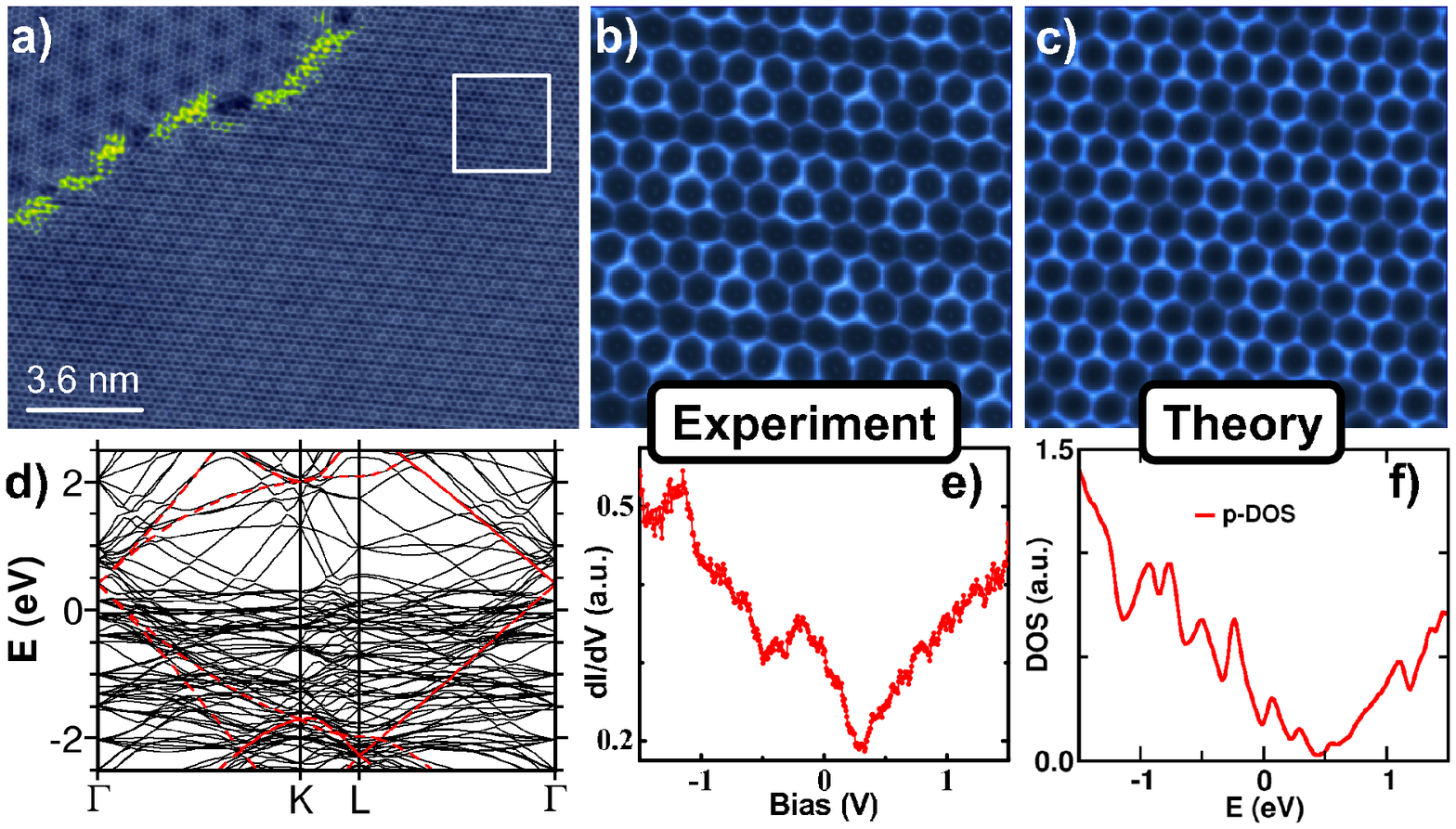}
\caption{\label{Fig1} (color online) a) STM image of the pristine graphene/Pt(111) surface showing two different moir\'e structures. b) Zoom in of the 3$\times$3 region highlighted in a). Sample bias: 50mV, tunneling current: 1.0nA for a) and b). STM data were measured and analyzed with WSXM \cite{Horcas'07}. c) Simulated STM image of 3$\times$3 graphene/Pt at V=+100 mV. d) Calculated band structure of the 3$\times$3 Graphene/Pt moir\'e (black lines) and pure graphene (red lines). Pure graphene bands have been shifted by +410 meV. e) STS measurement of the LDOS of pristine graphene on Pt(111).f) Theoretical DOS of the 3$\times$3 moir\'e projected on the p states of C atoms (s and d contributions are negligible in this energy range). }
\end{figure}

All the experimental results shown here are basically independent of the moir\'e  periodicity, thus, for the sake of simplicity and to facilitate a straightforward comparison with theory, we will restrict ourselves to the 3$\times$3 moir\'e (Fig. 1b), one of the most frequently found in the graphene/Pt(111) system \cite{Sutter'09}. Information about the local density of states (LDOS) of the sample was obtained with atomic precision by measuring at 6 K differential conductance (dI/dV) spectra in open feedback loop mode using the lock-in technique with frequency 2.3 kHz and ac modulation of 1 mV. Our dI/dV spectra measured in the pristine graphene surface show a clear dip at $\sim$ +300 mV accompanied by a {\em V}  shaped rise at both sides, see Fig. 1e. Recent scanning tunneling spectroscopy (STS) experiments measured on a partially covered graphene/Pt(111) surface, have reported  dI/dV spectra with a slight dip at +150 mV \cite{Levy'10}, which was related to the unoccupied surface state existing in Pt(111) \cite{Wiebe'05}. However, the clear {\em V} shape of our spectra around the dip and its location at +300 mV, coinciding with the Dirac point energy theoretically predicted \cite{Giovannetti'08} and estimated from photoemission experiments \cite{Sutter'09}, make us believe that the dip we observe at +300 mV is associated with the position of the Dirac point in the graphene/Pt(111) surface. Another piece of evidence comes from the results of our density functional theory (DFT) calculations \cite{SuppInfo} based on VASP \cite{Kresse96}. We have studied the 3$\times$3 moir\'e using an empirically corrected version of the PBE functional \cite{Perdew96} that includes the effects of van der Waals interactions \cite{Grimme06}. The calculated bands of the system are shown in Fig. 1d (in black). In the same figure we also include the bands of pure graphene (in red) shifted by +410 meV to make them overlap with their equivalent in the moir\'e. Consistently, the calculated DOS of Fig. 1f, which can be directly compared to the experiment in Fig. 1e, presents a V-shaped minimum at that same energy. Our theoretical results capture the structural properties of the graphene/Pt(111) system very well, in opposition to other methodologies that include dispersion interactions into standard DFT functionals \cite{Vanin10}. In particular, our calculated average distance between graphene and the uppermost Pt layer is 3.35 \AA, whereas LEED/LEEM experiments yield 3.30 \AA \cite{Sutter'09}.

Experimental STM images of graphene adsorbed on Pt(111) (Fig. 1b) can be understood in more depth with the help of theory. In our approach, we use a non-equilibrium Green's function formalism to evaluate the currents \cite{Blanco06}, using the OpenMX code \cite{openmx1} to map the hamiltonian into a local orbital basis, and an idealized Pt apex with a single dz${^2}$ orbital to represent the microscope tip. This model produces atomically resolved images (see Fig. 1c) in good agreement with the experimental results in Fig. 1b. Surprisingly, the brightest features in the calculated image correspond to C atoms that are lowest in the graphene sheet. Notice that in the 3$\times$3 moir\'e the lattice mismatch is very small, and the topographic corrugation, measured as the height difference between the lowest and highest C atoms, is only 0.02 \AA. In this case, the observed anticorrelation between the simulated image and the atomic topography implies that the STM corrugation is not a geometrical effect but instead it can be explained as a purely electronic effect. 

To atomically tailor the graphene layer adsorbed on Pt(111), we have generated point defects on it by irradiating the surface with 140 eV Ar${^+}$ ions, which are known to mainly produce single C vacancies on a graphite surface \cite{Ugeda'10}. Our LT-STM images show that the previously pristine graphene sheet presents now a number of almost identical bright features, associated with the amount of Ar${^+}$ ions reaching the surface, see Fig. 2. The exact shape of these features slightly depends on its location with respect to the moir\'e pattern, as shown in Figs. 2b-e. We see an elongated protrusion occupying an extension of two honeycomb lattice parameters, surrounded by a non trivial pattern of high LDOS intensity extending less than 1nm. In ideal graphene, the presence of a point defect should generate short-wavelength modulations of the LDOS with $(\sqrt{3}\times\sqrt{3})$ R30$^\circ$ (R3 in the following) periodicity due to intervalley scattering \cite{Ando'98}. In well-decoupled graphene layers, as epitaxial graphene on SiC or HOPG surfaces, such R3 patterns associated with atomic-size impurities have been observed by means of low bias STM images, which are a measure of the LDOS at E${_F}$, \cite{Kelly'96, Ugeda'10, Mallet'07,Rutter'07, Brihuega'08}. 2D Fourier transforms of our low bias STM images measured in samples with a high enough density of generated atomic vacancies, also allowed us to measure such R3 modulations and to estimate a Fermi wave-vector $k{ _F}\sim$ 0.5nm$^{-1}$, which is consistent with E${_D}$=+300meV and a Fermi velocity 10${^6}$m/s as recently measured by photoemission \cite{Sutter'09,SuppInfo}. 

\begin{figure}
\includegraphics[width=85mm]{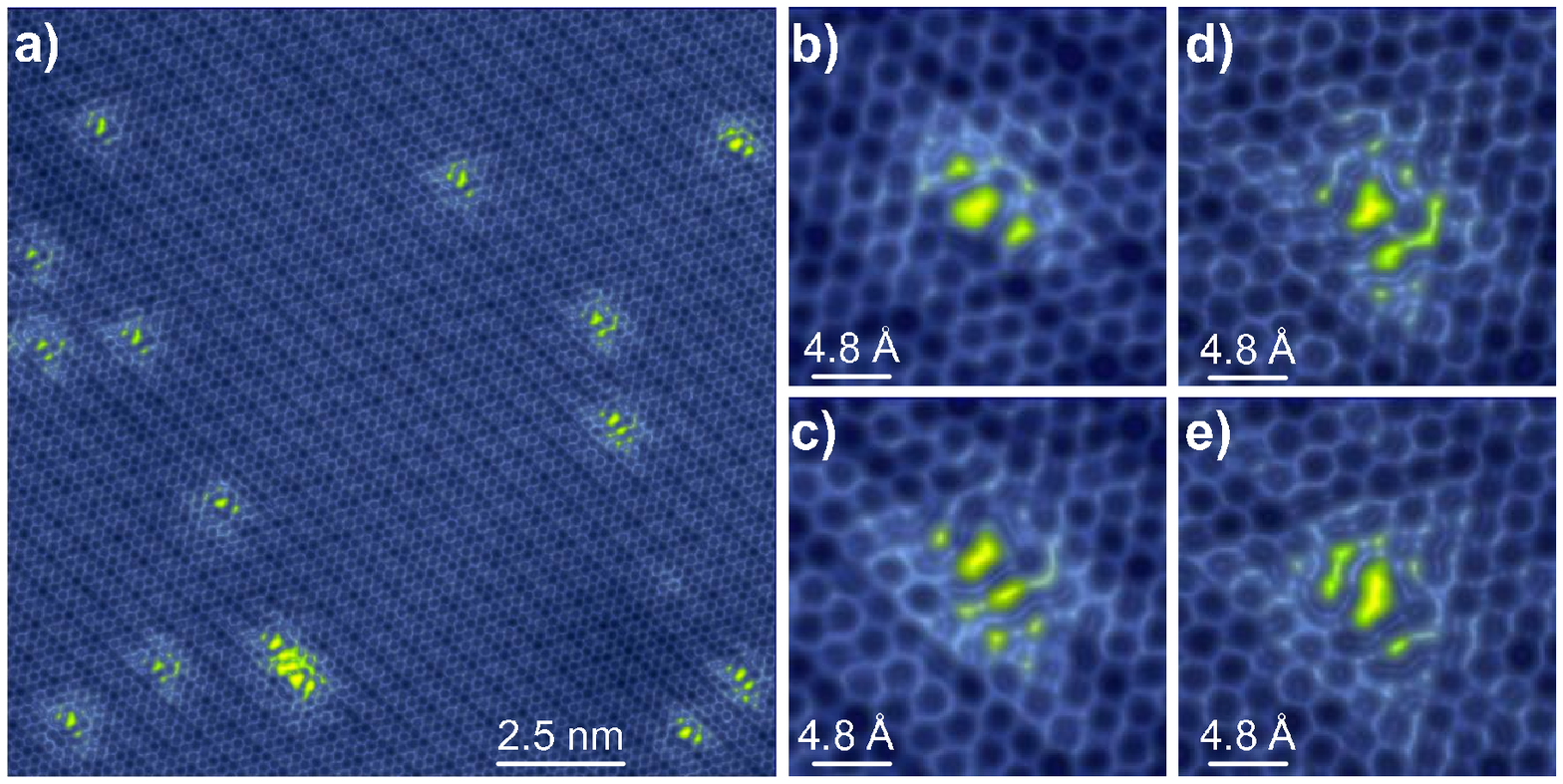}
\caption{\label{Fig2} (color online) a) STM topography, measured at 6 K, showing the graphene/Pt(111) surface after the Ar${^+}$ irradiation. Sample bias: -30mV, tunneling current: 0.8nA. b-e) Zoom-ins showing the small variation in the bright features shape for different moir\'e  positions. All bright features on the image can basically be identified with one of these 4 images. } 
\end{figure}

Contrary to the case of the graphite surface \cite{Ugeda'10}, we cannot directly identify point defects generated by Ar sputtering in the graphene/Pt(111) surface with the three fold patterns predicted for single C vacancies on a FSG layer \cite{Kelly'96}. Thus, we have to bring into play DFT calculations in order to unravel the nature of the atomic point defects observed in this system.
In our simulations, we use a supercell with 2$\times$2 units of the 3$\times$3 moir\'e and a single C atom removed \cite{SuppInfo}. For this structure, several adsorption sites are possible, depending on the relative location of the vacancy and the underlying Pt surface. We have studied some of these possibilities, finding that the best match to STM experiments occurs for structures which reconstruct similarly to single vacancies on isolated graphene sheets \cite{Yazyev07}. In the reconstruction of free standing graphene, two of the 3 undercoordinated C atoms surrounding the vacancy move closer to each other and become weakly bonded, forming a pentagon ring, and the third undercoordinated atom moves out of the graphene plane by only $\sim$ 0.1\AA. In graphene/Pt(111), the third atom and one of its neigbours move out of the plane and towards the Pt by $\sim$ 1 \AA, and form two new chemical bonds with Pt atoms of the surface (Figs. 3a and b). Besides, the average distance between the graphene sheet and the topmost Pt layer has decreased by $\sim$ 0.08 \AA. It is thus clear that the previously inert graphene layer has become very reactive. The strong interaction between the graphene with vacancy and the metal also results in a quench of the possible magnetic moment of the system, with the relaxed structure being non-magnetic. The simulated STM images present a pattern like that in Fig. 3c. The brightest features are a heart-shaped, elongated protrusion and a small, oval protrusion next to it. Another characteristic feature is a small dark area right next to the elongated protrusion. All these structures can be recognized in the experimental images (Fig. 3d). By comparing the simulated image with the position of the atoms (Fig. 3c) we can associate the heart-shaped feature to the region where the two weakly bonded C atoms and the vacancy itself are located, the dark region to the out-of-plane C atoms, and the small, oval feature to the other atom in the pentagon ring which belongs to the same sublattice as the weakly bonded pair. The modulation of the charge density revealed by the STM image decays fast. However, we do not try to compare it quantitatively with the experiment, since our unit cell is still small to model an isolated vacancy, and we believe the intensity decay can be affected by interference between neigbouring cells.

\begin{figure}
\includegraphics[width=85mm]{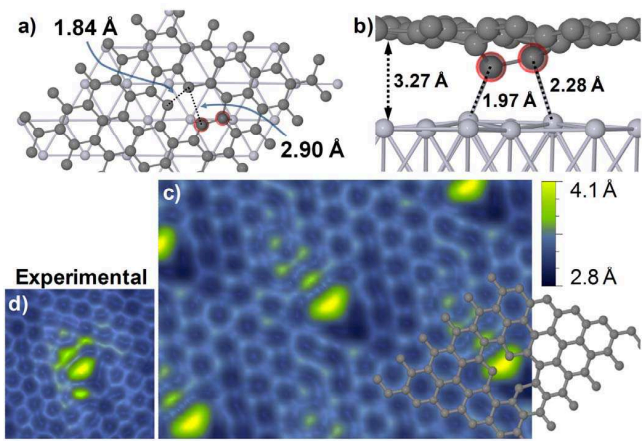}
\caption{\label{Fig3} (color online) Relaxed single carbon vacancy in a 6$\times$6 unit cell (4 cells of the 3$\times$3 graphene/Pt(111) moir\'e ), with the 2 C atoms that move towards the Pt highlighted: a) top view, b) side view, c) calculated constant current STM image at V = +100 mV with the ball and stick atomic atomic model superimposed. d) Experimental STM image of a vacancy from panel 2a.
}
\end{figure}

Introducing single C vacancies in an isolated graphene layer has a profound impact in its properties. According to many theoretical studies \cite{Pereira'06, Yazyev'08, Palacios'08}, they give rise to quasi-localized states at the Fermi level, which can be associated with the generation of local magnetic moments around the C vacancies and produce a strong reduction of charge carriers' mobility. Such theoretical expectations were recently confirmed by some of us by STS experiments on C vacancies on the graphite surface, which revealed the presence of a very sharp resonance at the Fermi energy extending more than 3 nm away from each single C vacancy \cite{Ugeda'10}. It is thus crucial to understand how the coupling with the metallic substrate will affect the properties of such C vacancies.

Our STS measurements on C vacancies in graphene/Pt(111) show a strong increase of the LDOS starting at the Dirac point and reaching a maximum around +500 meV. As it can be seen in Fig. 4a, this is reflected in our dI/dV spectra as a broad electronic resonance centered at +500 meV with a FWHM $\sim$150 meV. dI/dV spectra acquired on C vacancies located in different positions of the 3$\times$3 moir\'e, show that all C vacancies present this broad electronic resonance at +500 meV independently of its position inside the moir\'e, which points to a small influence of the moir\'e superstructure on its properties. The only influence detected was the variation of the resonance height for C vacancies on different moir\'e  positions, see Figs. 4b-c. Analogous results were also obtained for different moir\'e patterns. Since the electronic resonance associated with any vacancy lays at a fixed energy of +500 meV, it is possible to map its spatial extension by conductance images at +500 mV. Figure 4c shows one of such conductance images measured on a 15$\times$15 nm${^2}$ region where 11 single C vacancies have been generated by Ar${^+}$ irradiation. As it can be observed, the localized state only extends 0.6 nm away from the center of the vacancies. The conductance map also reflects the resonance height variation found for vacancies in different moir\'e locations. Our calculations show that this peak is mainly caused by the ${\pi}$-electrons of the two weakly bound atoms in the graphene reconstruction. 

\begin{figure}
\includegraphics[width=75mm]{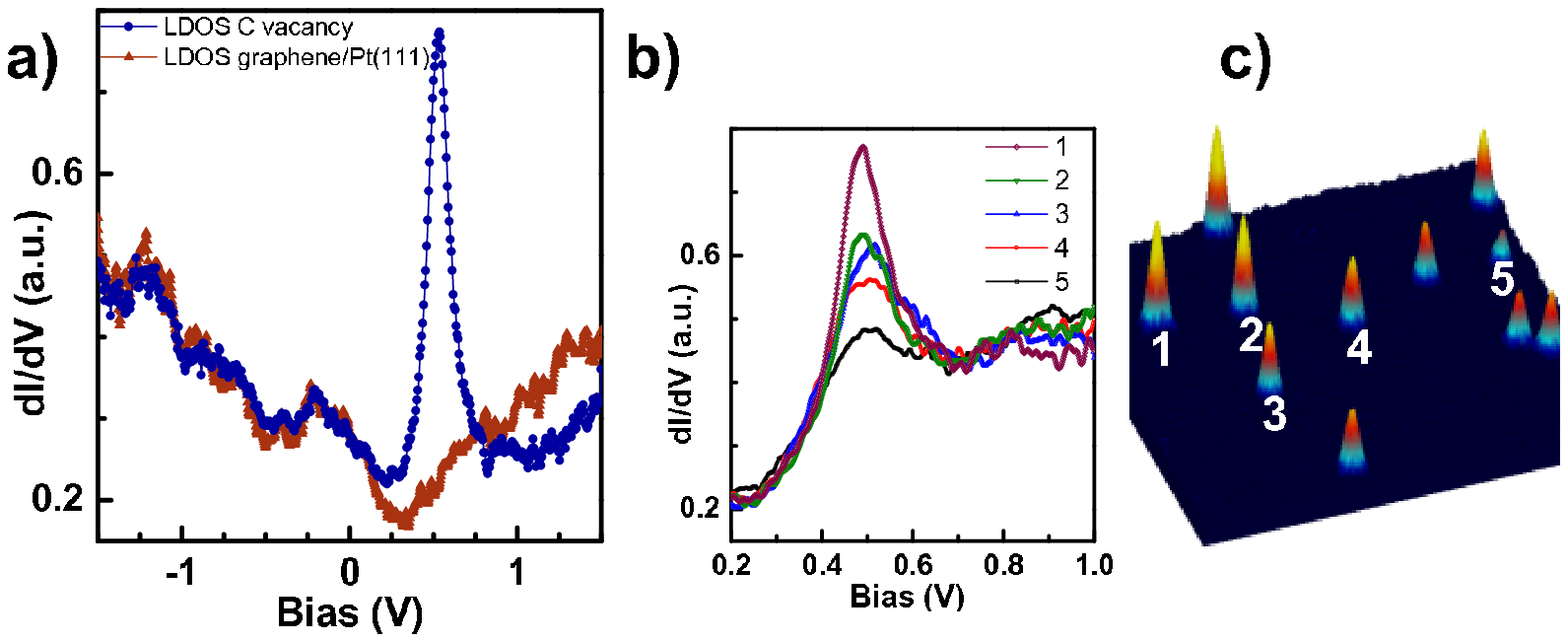}
\caption{\label{Fig4} (color online) a) STS measurements of the LDOS on a C vacancy (blue) and on pristine graphene/Pt(111) (red). dI/dV were taken consecutively at 6 K with the same microscopic tip. b) dI/dV spectra measured on different C vacancies, numbered 1 to 5. c) dI/dV map, measured at a sample bias of +500mV. 1 to 5 point to same vacancies as panel b). 
}
\end{figure}

Thus, our results show that the electronic properties of C vacancies in a graphene layer on top of a metal strongly differ from the ones found for C vacancies in a well decoupled graphene layer as it is the graphite surface\cite{Ugeda'10}. As shown by our DFT calculations these differences can be explained as due to the increase of the interaction with the metallic surface after the C vacancy formation. 

In summary, our findings demonstrate that even in weakly coupled graphene/metal systems the presence of the metal has to be seriously taken into account in order to controllably tune graphene properties by locally modifying its structure. We have shown that the interaction with the metal strongly increases when single C vacancies are introduced in the graphene layer, discarding the appealing possibility of inducing magnetic properties on this graphene system by removing C atoms.
\begin{acknowledgments}
This work is supported by MAT2008-02929-NAN, MAT2010-14902, MAT2008-02939-E, CSD2010-00024 (MICINN, Spain), by PERG05-GA-2009-24209 (EU) and S2009/MAT-1467 (Comunidad de Madrid, Spain). I.B and P.P. were supported by the Ram\'on y Cajal program. Computer time provided by the Spanish Supercomputing Network (RES), and by the Centro de Supercomputacion y Visualizacion de Madrid (CeSViMa).

 \end{acknowledgments}


\begin{thebibliography}{42}

\bibitem{Wallace'47}	P. R. Wallace, Physical Review {\bf 71}, 622 (1947).
\bibitem{Novoselov'04}	K. S. Novoselov {\em et al.}, Science {\bf 306}, 666 (2004).
\bibitem{Novoselov'05b}	K. S. Novoselov {\em et al.}, Nature {\bf 438}, 197 (2005b).
\bibitem{Zhang'05}	Y. B. Zhang {\em et al.}, Nature {\bf 438}, 201 (2005).
\bibitem{Miller'09}	D. L. Miller {\em et al.}, Science {\bf 324}, 924 (2009).
\bibitem{Castro Neto'09}	A. H. Castro Neto {\em et al.}, Rev. Mod. Phys. {\bf 81}, 109 (2009).
\bibitem{Geim'09}	A. K. Geim, Science {\bf 324}, 1530 (2009).
\bibitem{N'Diaye'06}	A. T. N'Diaye {\em et al.}, Phys. Rev. Lett. {\bf 97} 215501 (2006).
\bibitem{de Parga'08}	A. L. V. de Parga {\em et al.}, Phys. Rev. Lett. {\bf 100} 056807 (2008).
\bibitem{Wintterlin'09}	J. Wintterlin, $\&$ M. L. Bocquet, Surf. Sci. {\bf 603}, 1841 (2009).
\bibitem{Kim'09}	K. S. Kim {\em et al.}, Nature {\bf 457}, 706 (2009).
\bibitem{Li'09}	X. S. Li {\em et al.}, Science {\bf 324}, 1312 (2009).
\bibitem{Giovannetti'08}	G. Giovannetti {\em et al.}, Phys. Rev. Lett. {\bf 101} 026803 (2008).
\bibitem{Preobrajenski'08}	A. B. Preobrajenski {\em et al.}, Phys. Rev. B {\bf 78} 073401 (2008).
\bibitem{Gruneis'08}	A. Gruneis, $\&$ D. V. Vyalikh, Phys. Rev. B {\bf 77} 193401 (2008).
\bibitem{Pletikosic'09}	I. Pletikosic {\em et al.}, Phys. Rev. Lett. {\bf 102} 056808 (2009).
\bibitem{Sutter'09}	P. Sutter, J. T. Sadowski, $\&$ E. Sutter, Phys. Rev. B {\bf 80} 245411 (2009).
\bibitem{Land'92}	T. A. Land {\em et al.}, Surf. Sci. {\bf 264} 261 (1992).
\bibitem{Balog'10}	R. Balog {\em et al.}, Nat. Mater. {\bf 9}, 315 (2010).
\bibitem{Enachescu'99}	M. Enachescu{\em et al.}, Phys. Rev. B {\bf 60}, 16913 (1999).
\bibitem{Otero'10} G. Otero {\em et al.} Phys. Rev. Lett. {\bf 105}, 216102 (2010).
\bibitem{Miguel'11} M. M. Ugeda, doctoral thesis, UAM (to be published).

\bibitem{Levy'10}	N. Levy {\em et al.}, Science {\bf 329}, 544 (2010).
\bibitem{Wiebe'05}	J. Wiebe {\em et al.}, Phys. Rev. B {\bf 72} 193406 (2005).
\bibitem{SuppInfo} See supplementary material at http://link.aps.org/supplemental/XXX.
\bibitem{Kresse96} G. Kresse $\&$ J. Furthmüller, Phys. Rev. B 54, 11169 (1996); G. Kresse $\&$ D. Joubert, Phys. Rev. B 59, 1758 (1999).
\bibitem{Perdew96} J.P. Perdew, K. Burke, $\&$ M. Ernzerhof, Phys. Rev. Lett. 77, 3865 (1996).
\bibitem{Grimme06} S. Grimme, J. Comp. Chem. {\bf 27}, 1787 (2006).
\bibitem{Vanin10} M Vanin {\em et al.}, Phys. Rev. B {\bf 81}, 081408(R) (2010).
\bibitem{Blanco06} J.M. Blanco, F. Flores, $\&$ R. Perez, Prog. Surf. Sci. {\bf 81}, 403 (2006).
\bibitem{openmx1} http://www.openmx-square.org; T. Ozaki, Phys. Rev. B, {\bf 67} (2003), 155108).
\bibitem{Ugeda'10}	M. M. Ugeda {\em et al.}, Phys. Rev. Lett. {\bf 104} 096804 (2010).
\bibitem{Ando'98}	T. Ando, T. Nakanishi, and R. Saito, J. Phys. Soc. Jpn. {\bf 67}, 2857 (1998).
\bibitem{Kelly'96}	K. F. Kelly {\em et al.}, Science {\bf 273}, 1371 (1996).
\bibitem{Mallet'07}	P. Mallet {\em et al.}, Phys. Rev. B {\bf 76} (2007).
\bibitem{Rutter'07}	G. M. Rutter {\em et al.}, Science {\bf 317}, 219 (2007).
\bibitem{Brihuega'08}	I. Brihuega {\em et al.}, Phys. Rev. Lett. {\bf 101} 206802 (2008).
\bibitem{Yazyev07} O. V. Yazyev $\&$ L. Helm, Phys. Rev. b {\bf 75}, 125408 (2007).
\bibitem{Pereira'06}    V. M. Pereira {\em et al.}, Phys. Rev. Lett. {\bf 96} 036801 (2006).
\bibitem{Yazyev'08}	O. V. Yazyev, Phys. Rev. Lett. {\bf 101} 037203 (2008).
\bibitem{Palacios'08}	J. J. Palacios, J. Fernandez-Rossier, and L. Brey, Phys. Rev. B {\bf 77} 195428 (2008).
\bibitem{Horcas'07} I. Horcas {\em et al.}, Rev. Sci. Instrum. {\bf 78} 013705 (2007).


\end{thebibliography}
\end{document}